\input{aipcheck}

\documentclass[
    ,final            
  ]
  {aipproc}

\layoutstyle{6x9}

\begin{document}

\title{Parameter identification using the Hilbert transform}

\author{Andrew Allison}{address={Centre for Biomedical Engineering (CBME)
and \\ Department of Electrical and Electronic Engineering, University of Adelaide, SA 5005, Australia.}}

\author{Derek Abbott}{address={Centre for Biomedical Engineering (CBME)
and \\ Department of Electrical and Electronic Engineering, University of Adelaide, SA 5005, Australia.}}

\begin{abstract}
Many physical systems can be adequately modelled using a second order
approximation. The problem of plant identification reduces to the
problem of estimating the position of a single pair of complex
conjugate poles. 

One approach to the problem is to apply the method of least squares to
the time domain data. This type of computation is
best carried out in "batch" mode and applies to an entire data
set. Another approach would be to design an adaptive filter and to use
autoregressive, AR, techniques. This would be well suited to
continuous real-time data and could track slow changes on the
underlying plant.

I this paper we present a very fast but approximate technique for the
estimation of the position of a single pair of complex conjugate
poles, using the Hilbert transform to reconstruct the analytic signal.
\end{abstract}

\maketitle


\section*{introduction}

In the theory of control, it is most common for physical systems to be
mathematically modelled using coupled linear systems of ordinary
differential equations\cite{levine_1996}. When these equations are
transformed using integral transforms, such as those of Laplace or
Fourier, then the physical systems are modelled using finite rational
polynomials in an auxiliary variable, $s = j \omega$:
\begin{equation}
   H(s) = \frac{\mbox{output}(s)}{\mbox{input(s)}} = \frac{P(s)}{Q(s)}~.
   \label{eq:rational_polynomial}
\end{equation}
The zeros of the polynomial, $Q(s)$, are called ``poles'' and
correspond to responses that have finite output for zero input. These
are called "modes." It is very common for one mode to dominate the
response of the whole system. It is also common for this mode to be of
a damped oscillatory type, corresponding to a single pair of complex
conjugate poles. This can occur whenever the potential energy function
of the system has a local minimum \cite{goldstein_1950} In this case,
we can approximate a large complicated system, with many poles and
zeros, by a simple second-order system with a single pair of complex
conjugate poles.  This is called the second-order approximation. Many
mechanical or electrical systems can be realistically modelled using
the second-order approximation.
We can write:
\begin{equation}
   H(s) \approx \frac{a_2 s^2 + a_1 s + a_0 }
               { s^2 + 2 \alpha s + {\omega_0}^2 }~.
   \label{eq:canonical_second_order}
\end{equation}

If we wish to model the behaviour of a real physical system, using an
approximate second-order model, then it is necessary for us to estimate the
position of the pole pair. This could be done in the frequency domain,
by exciting the system with a sinusoidal source and then measuring the
magnitude and phase of the response at different frequencies, but
this is often not practical. There are situations when the only practical
sources are step functions, $1(t)$ or impulses $\delta(t)$. We can
excite the system with steps or impulses and then sample the response in
the time domain. The impulse response of a second-order system will
generally be of the form:
\begin{equation}
   y(t) = A e^{- \alpha t} \cos \left( \omega_d t \right) +
          B e^{- \alpha t} \sin \left( \omega_d t \right)
   \label{eq:second-order_time_domain}
\end{equation}
where ${\omega_d}^2 = {\omega_0}^2 - {\alpha}^2$.

The problem of plant identification then becomes
equivalent to asking: How do we estimate the position of the pair of
complex-conjugate poles if the only data at our disposal is a set of
time-domain samples of the response of the system to steps or impulses.

As a possible illustration, we could imagine that we strike a bell
with a hammer and then record the sound as it gradually decays.  We
want to estimate the damped frequency of oscillation, $\omega_d$, and
the damping coefficient, $\alpha$, using {\it only} the data from our sound
recording.

If we knew something about the distribution of the errors of
measurement then we could apply the method of maximum likelihood to
estimate the parameters , $\omega_d$ and $\alpha$. If the errors were
known to be the result of a very large number of uncorrelated random
effects then we could apply the Central Limit Theorem and we could
assume that the errors had a Gaussian distribution.  The problem of
plant identification would reduce to a non-linear least-squares
estimation problem \cite{press_1995}. The difficulty with this approach is that the
resulting equations would be non-linear and would have to be solved
iteratively, using a numerical method such as gradient descent. A
further weakness of this approach is that it would be an {\it exact}
solution to an {\it approximation} of the real problem. It would be
far more reasonable to have a quick but approximate solution to the
approximate, second-order, problem. This would tell us most of what we
need to know without having to waste a lot of effort.

In this paper, we present a fast, but approximate, algorithm for the
estimation of the position of a complex conjugate pair of poles on the
$s$ plane. We use a discrete approximation to the  Hilbert Transform , to reconstruct
the complex analytic signal form the sampled real time signal.  The
analytic signal has a complex-exponential form. We apply very simple
statistical techniques to the analytic signal in order to obtain the
required parameters.  This approach is an alternative to the more
conventional, non-linear least squares or Autoregressive, AR,
approaches to the problem.

The problem of parameter identification, for a freely vibrating system
has been studied by Feldman \cite{feldman_1997} who used the Hilbert
Transform to provide information about instantaneous amplitude and
phase of a signal. The method that we present here is more simple, and
limited, than the approach used by Feldman.

\section*{The analytic signal of the second order response}

The response described in Equation \ref{eq:second-order_time_domain} is equivalent to
\begin{equation}
   y(t) = C e^{- \alpha t} \cos \left( \omega_d t - \phi \right)
   \label{eq:real_time_signal}
\end{equation}
where $C = \sqrt{A^2 +B^2}$, $\cos(\phi) = A / \sqrt{A^2+B^2}$ and
$\sin(\phi) = B / \sqrt{A^2+B^2}$.  This type of function will apply
whenever the input to the system is zero. If the input is a finite sum
of step and impulse function then the input will be zero for most of
the time. There will be abrupt changes in $C$ and $\phi$ but the
parameters, $\alpha$ and $\omega_d$ will be constant as long as the
structure of the plant is maintained.

The immediate aim is to reconstruct the analytic signal.  The Hilbert
transform \cite{bracewell_1965,hahn_1996,ersoy1997} is a standard
technique for achieving this.  The Hilbert transform of $u(t)$ is
defined as:
\begin{equation}
   v(s) = \frac{1}{\pi} \int_{-\infty}^{+\infty} \frac{u(t)}{s-t} dt~.
   \label{eq:hilbert_transform}
\end{equation}
It has the important property that it defines the relationship between
the real and imaginary parts of a complex analytic function. 
If we have an analytic function:
\begin{equation}
   \Phi(t) = u(t) + j v(t)
   \label{eq:analytic}
\end{equation}
with real and imaginary parts $u(t)$ and $v(t)$ then the relation for $v(s)$ is given by
Equation \ref{eq:hilbert_transform}.
Bedrosian's theorem tells us that the Hilbert transform of $a(t)
\cos(\omega t)$ is $a(t) \sin(\omega t)$. If we apply Bedrosian's
\cite{hahn_1996}theorem and the shifting property to Equation
\ref{eq:real_time_signal} then we find that the analytic signal is:
\begin{eqnarray}
   y_2(t) & = & C e^{- \alpha t} \cos \left( \omega_d t - \phi \right) + j C e^{- \alpha t} \sin \left( \omega_d t - \phi \right) \\
        & = & C e^{-j \phi } e^{\left( - \alpha + j \omega_d \right) t}~.
   \label{eq:analytic signal}
\end{eqnarray}
This analytic signal is a pure exponential function and can
essentially be ``unwrapped'' using the $\log()$ function. We can write:
\begin{equation}
   \log \left(  y_2(t) \right) = \log(C) - j \phi +  \left( - \alpha + j \omega_d \right) t~.  
   \label{eq:log_analytic}
\end{equation}
If we sample this analytic signal at intervals of $T_s$ then we can
numerically calculate the slope of Equation \ref{eq:log_analytic} to get:
\begin{equation}
   -\alpha + j \omega_d =  \left(  \frac{ \log \left( y_2 \left( t + T_s \right) \right) - 
       \log \left( y_2 \left( t \right) \right) }
       {T_s} \right)~.  
   \label{eq:slopes}
\end{equation}
This allows us to directly estimate the parameters, $\alpha$ and
$\omega_d$.  We note that the use of the difference operation has
removed all reference to $C$ and to $\phi$. This means that the method
is not sensitive to the initial conditions that apply immediately
after the shocks that occur then the impulses and steps are fed into
the system. Our only requirement is that the input to the system is
zero for most of the time.

The Matlab code required to implement this algorithm is very short and simple:
\begin{verbatim}
% reconstruct the analytic signal, y2
y2 = hilbert( y );
% take the natural logarithm, log(y2)
L = log(y2) ;
% unwrap the phase of the log(y2)
L = real(L) + j*unwrap(imag(L));
% estimate the differences of the log of y2
D = diff(L) ;
\end{verbatim}

\section*{A simple statistical technique}

Equation \ref{eq:slopes} suggests that we should be able to precisely
estimate the required parameters. There are a few practical problems with 
the direct application of Equation  \ref{eq:slopes}:

\begin{itemize}

   \item~The numerical calculation of the Hilbert transform relies on
   the Fast Fourier Transform and there are limitations imposed by the
   finite number of samples. These includes the ``Gibbs effect,'' due to
   the finite length of the data set.

   \item~Real samples from a physical process will be subject to noise
   and errors of measurement. The differencing operation tends to
   magnify the effect of noise.

   \item~There would be a number of outliers caused by the ``shocks''
   of the steps or impulses. Some measurements will not be reliable.

\end{itemize}

The authors have found that the median is a very robust measure of
location and is less subject to the influence of the outlying values
than the arithmetic mean. The Matlab code for this is very simple: 

\begin{verbatim}
% calculate the real and imaginary parts of the differences
d_mag   = real(D) ;
d_phase = imag(D) ;
% calculate the median rates of change
mid_re_slope = median(d_mag) ;
mid_im_slope = median(d_phase) ;
\end{verbatim}

\section*{Some results}

\begin{figure}[h!]
\includegraphics*[width=0.9\textwidth]{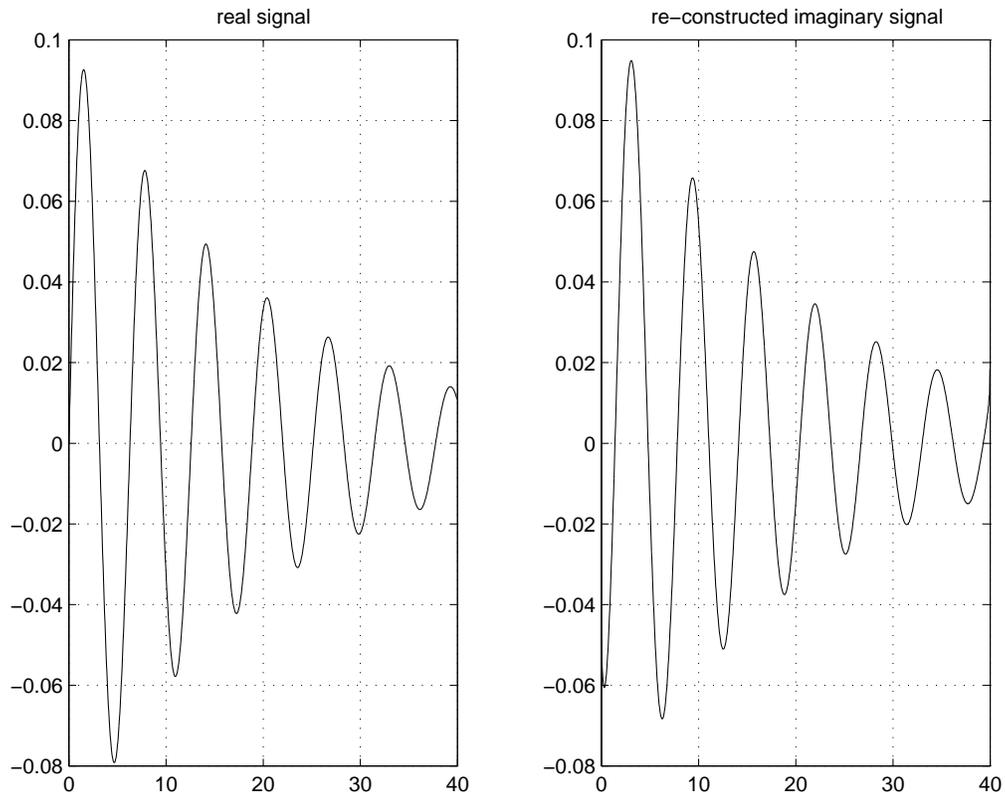}
\caption{ The original sampled signal is shown on the left. The signal on
the right is an estimate of the complex part of the analytic signal
which was reconstructed using the Hilbert Transform.  There is
significant error in the reconstructed signal near the sample
boundaries, due to the Gibbs effect. This is the result of fact that
Matlab uses the FFT to calculate the Hilbert transform.}
\label{fig:fig1}
\end{figure}

A second order system was simulated, using known parameters, and the
parameters were then estimated using the new algorithm. The
reconstructed analytic signal is shown in Figure \ref{fig:fig1}.

\begin{figure}[h!]
\includegraphics*[width=0.9\textwidth]{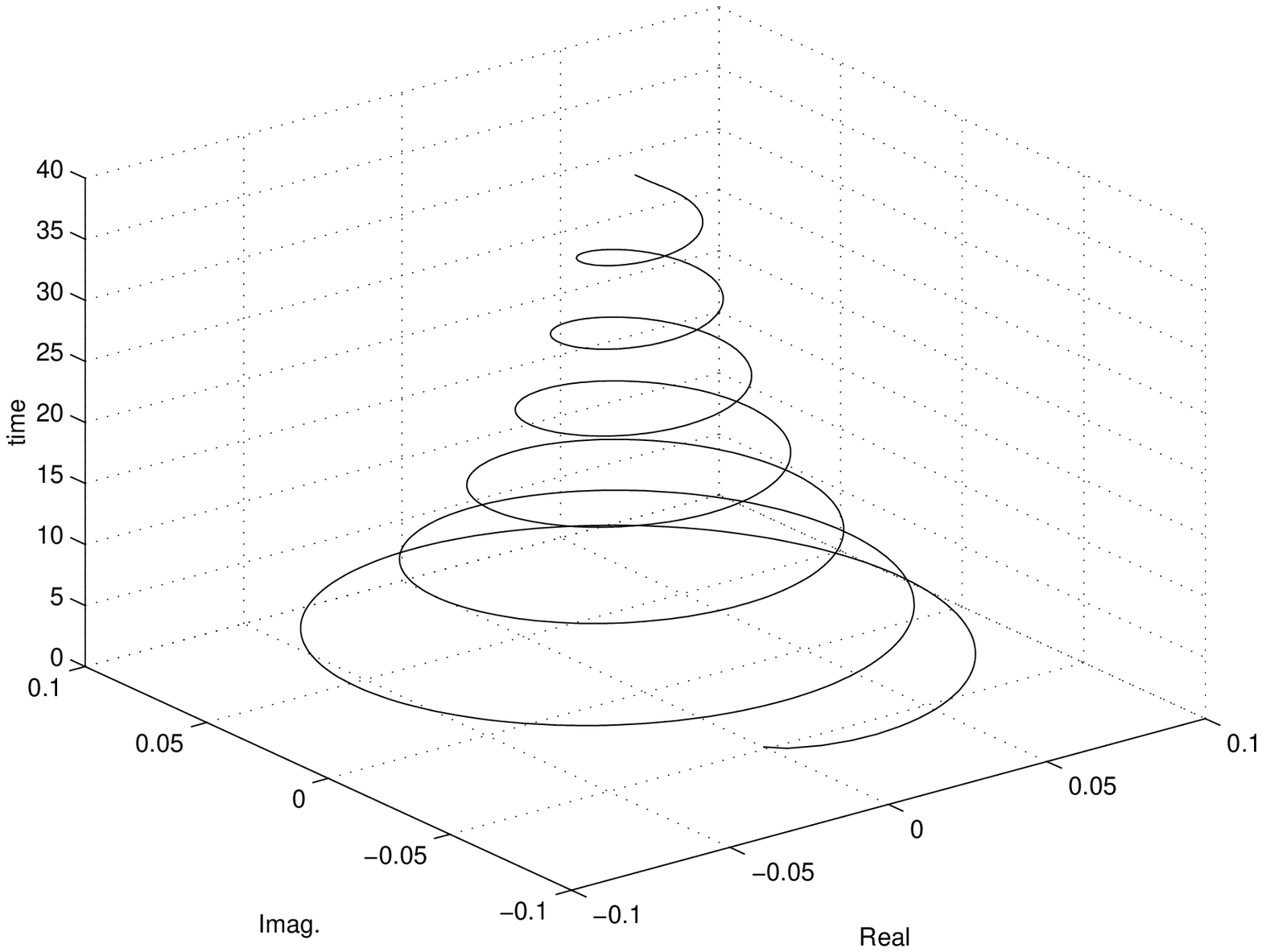}
\caption{This is a 3D plot of the reconstructed analytic signal. The XY
plane and all planes parallel to it represent the complex field that
contains the analytic signal. The vertical, or Z, axis represents
time. The general appearance of damped oscillation is unmistakable.}
\label{fig:fig2}
\end{figure}

This same data can be represented in three dimensions. This is shown
in Figure \ref{fig:fig2}.

The logarithmic slopes were estimated and the parameters were calculated.

The quality factor of the simulated system was $Q=10$ and so the
theoretical value of the damping constant was $\alpha=0.05$. The
estimated value was $\alpha_{\mbox{est}} = 0.0472$. The characteristic
frequency of the simulated plant was $\omega_d = 1.0$ and the estimate
was $\omega_{\mbox{est}} = 0.9948$.

There was a $-5.5 \%$ error in the estimate of the damping constant,
$\alpha$ and there was a $-0.4\%$ error in the estimate of the damped
natural angular frequency of oscillation, $\omega_d$.

\section*{Summary and limitations }

The method does work as an approximation and could estimate $\alpha$
to within $-5.5 \%$ and $\omega_d$ to within $-0.4\%$.  The error in
the estimate of the damping constant, $\alpha$ and there was a
$-0.4\%$ error in the estimate of the damped natural angular frequency
of oscillation, $\omega_d$.  There are some notable problems with the
method:
\begin{itemize}

\item~It is sensitive to relative time scale of the constants in question
and the width of the sampling window.

\item~It suffers from boundary effects due to the Gibbs' phenomenon. 

\item~The method is also numerically unstable for large data sets with large
values of time, $t$, since $exp(-\alpha t)$ can cause Matlab to
underflow.

\end{itemize}
Our simulations suggest that this is a fast but approximate technique
for the estimation of the position of a single pair of complex
conjugate poles but it is sensitive to a number of factors which may
limit its practical application.


\bibliography{hilbert}
\bibliographystyle{aipproc} 

\end{document}